\documentclass[12pt]{iopart}
\usepackage{iopams}
\usepackage{epsfig}
\usepackage{epstopdf}
\usepackage{epsf}
\usepackage{subfigure}
\usepackage{graphicx}
\usepackage{bm}
\usepackage{float}
\usepackage{sidecap}
\usepackage{wrapfig}
\usepackage{wasysym}
\usepackage{amssymb}
\usepackage{color}

\newcommand{\bu}{\boldsymbol{u}}

\newcommand{\bx}{\boldsymbol{x}}
\newcommand{\bk}{\boldsymbol{k}}

\newcommand{\Xp}{\boldsymbol{X}}

\newcommand{\de}{\mathrm{d}}
\newcommand{\dd}[2]{\frac{\de{#1}}{\de{#2}}}

\begin{document}
\title[Lagrangian statistics under decimation]{Lagrangian Statistics for Navier-Stokes Turbulence under Fourier-mode reduction: Fractal and Homogeneous Decimations \footnote{Postprint version of the article published on New. J. Phys. 18 113047 (2016).}}
\author{Michele Buzzicotti$^1$, Akshay Bhatnagar$^2$, Luca Biferale$^1$, Alessandra 
S. Lanotte$^3$ \footnote{Author to whom all correspondence should be addressed.}, 
and Samriddhi Sankar Ray$^4$}
\address{$^1$ Dept. Physics and INFN, University of Rome ``Tor
  Vergata'', Via della Ricerca Scientifica 1, 00133, Rome, Italy.\\
$^2$ Nordita, KTH Royal Institute of Technology and Stockholm
  University, Roslagstullsbacken 23, 10691 Stockholm, Sweden.\\
$^3$ ISAC-CNR and INFN Sez. Lecce, 73100, Lecce, Italy.\\
$^4$ International Centre for Theoretical Sciences, Tata
  Institute of Fundamental Research, Bangalore 560089, India. \\}
\ead{a.lanotte@isac.cnr.it}
\begin{abstract}
We study small-scale and high-frequency turbulent fluctuations in
three-dimensional flows under Fourier-mode reduction. The
Navier-Stokes equations are evolved on a restricted set of modes,
obtained as a projection on a {\it fractal} or {\it homogeneous}
Fourier set.  We find a strong sensitivity (reduction) of the
high-frequency variability of the Lagrangian velocity fluctuations on
the degree of mode decimation, similarly to what is already reported
for Eulerian statistics.  This is quantified by a tendency towards a
quasi-Gaussian statistics, i.e., to a reduction of intermittency, at
all scales and frequencies. This can be attributed to a strong
depletion of vortex filaments and of the vortex stretching mechanism.
Nevertheless, we found that Eulerian and Lagrangian ensembles are
still connected by a dimensional {\it bridge-relation} which is
independent of the degree of Fourier-mode decimation.
\end{abstract}
\section{Introduction}

Turbulence is considered a key fundamental and applied problem
\cite{frischbook,pope}. Turbulent flows are distinguished in nature
and in the laboratories by the stirring mechanisms and the boundary
conditions. Both can be strongly anisotropic, non-homogeneous, and
non-stationary, leading to very different realizations for the mean
quantities and large-scale velocity configurations. In spite of this
large variety, we know that the central feature of all turbulent flows
stems from the non-linear terms which are able to transfer to all
scales the energy injected by the stirring mechanisms. The non-linear
terms are invariant under translation, rotation and mirror
symmetries. This is why isotropic, homogeneous and mirror symmetric
turbulence is considered a paradigmatic problem for fundamental and
applied studies \cite{frischbook}.

It is an empirical observation that, in three-dimensional turbulence, 
energy tends to be transferred from large to small scales
intermittently, i.e., producing larger and larger non-Gaussian
fluctuations by increasing the Reynolds number (the relative intensity
of non-linear versus linear terms in the equations). This is
accompanied by the development of anomalous power law scaling for the
moments of the velocity increments in the inertial range, i.e., at
scales much smaller (larger) than those where the forcing (viscous)
term acts. Intermittency of three-dimensional turbulence is not yet
fully understood. We cannot connect it to the equation of
motion. Neither can we predict its degree of universality, nor the key
dynamical and topological ingredients of its origins. For example,
two-dimensional turbulent flows are non intermittent with
quasi-Gaussian statistics in the inverse cascade regime
\cite{BCV2000}.
 
In the past, the Navier-Stokes equations (NSE) restricted on a sub-set
of Fourier modes have been numerically investigated to gain
information about the nature of anomalous scaling, its dependency on
the Reynolds number \cite{GLR1996,MPPZ1996,DeLillo2007}, and the
effect of local versus non local dynamics on the degree of
intermittency see e.g. \cite{DLN2001}. More recently, a new decimation
protocol has been proposed to ask further questions about the origin
of intermittency in the NSE \cite{frisch2012,PRL2015}. The idea is
again to selectively remove degrees of freedom in the Fourier space,
but now implemented in a way to preserve the same conserved quantities
and the same symmetries of the original undecimated NSE. By studying
the impact of such removal on the flow statistics (in particular, on
the intermittent behavior through changing of the projection
protocol), a better understanding about the degree of universality and
sensitivity of anomalous scaling in turbulence can be achieved.

In this paper, we follow this route further by investigating for the first time the
effects of Fourier-mode reduction on the evolution of Lagrangian
tracers in turbulence and thus also assessing temporal
intermittency. It is well known that Lagrangian particles get strong
feedback from the presence of small-scale intense vortex filaments
\cite{bode2001,bif2005,pumir2016}. Studying Lagrangian intermittency
under Fourier-mode reduction is therefore a direct way to quantify the
robustness of vortex stretching and small-scale vorticity production
mechanisms by  changing the active degrees of freedom in the dynamical
evolution.

We perform a series of direct numerical simulations (DNS) of the
three-dimensional NSE by restricting the dynamical evolution on a
prescribed quenched set of Fourier modes, and by varying the Reynolds
number. We investigate here the case when such a set of modes is a
{\it fractal} or {\it homogeneous} subset of the whole Fourier space.
Both these decimation methods belong to the class of spectral
    tools aiming at solving Navier-Stokes dynamics on a reduced set of
    wave numbers. The main goals of our theoretical and numerical work
    are: (i) understanding the impact of mode reduction on the
    Lagrangian statistics, and (ii) understanding the robustness of
    Eulerian-Lagrangian bridge relation at changing the scaling
    properties of the flow.\\
The effects on the Eulerian
statistics induced by the restriction of the dynamics on a fractal set
for two- and three-dimensional incompressible turbulence
\cite{frisch2012,PRL2015,rayreview,EPJE2016}, as well as on the
one-dimensional Burgers equations \cite{PRE2016}, have already been
reported.

In this paper we extend the previous findings on Eulerian
intermittency by considering the case of the homogeneous
mode-reduction and by studying its effects on the Lagrangian
statistics.  We find that homogeneous Fourier-mode decimation is a
quasi-singular perturbation for the Lagrangian scaling properties,
similar to what is seen for the Eulerian ones. Notwithstanding this
fact, we also find that Eulerian and Lagrangian statistics remain
strongly correlated, such that the bridge-relation empirically
observed for the original undecimated Navier-Stokes equations still
holds in the presence of Fourier-mode reduction.

This paper is organized as follows. In Section~\ref{sec:equations}, we
introduce the model equations for the Eulerian and Lagrangian
dynamics, as well as the decimation protocols; we also define the
set-up of the numerical experiments performed, together with the
relevant parameters. In section~\ref{sec:results} we separately
discuss the main results for the velocity field in terms of the
Eulerian and Lagrangian statistical properties; while in
Section~\ref{sec:bridge} we combine them together by quantitatively
assessing the validity of the bridge relation
\cite{Borgas1993,PRL2004,Schmitt2006,PRL2008,JOT2013,Boff2002}.
Summary and discussions are contained in the last section.

\section{Model equations for the Eulerian and Lagrangian dynamics}
\label{sec:equations}

\subsection{The  decimated equations of motion}

Let us define $\bu(\bx,t)$ and $\hat{\bu}(\bk,t)$ as the real and
Fourier space representations of the velocity field, respectively, in dimension
$D=3$.  We start by considering the Navier-Stokes
equations for the incompressible flow with unit density:
\begin{equation}
\label{eq:NS}
\partial_t {\bu} = - {\bf \nabla}p - (\bu \cdot {\bf \nabla}) \bu +
\nu \,\nabla^2 \bu + {\bf F}\, \end{equation} where $p$ is the
pressure and $\nu$ is the kinematic viscosity. ${\bf F}$ is a
homogeneous and isotropic forcing which drives the system to a
non-equilibrium statistically steady state. Decimation on a generic
sub-set of Fourier modes is accomplished by using a generalized
Galerkin projector, ${\cal P}$, which acts on the velocity field as
follows:
\begin{equation}
\label{eq:decimOper}
{\bf v}(\bx,t)= {\cal P} \, {\bf u}(\bx,t)=\hspace{-1mm}
\sum_{{\bk}}\hspace{-1mm} e^{i {\bk \cdot \bx}}\,\gamma_{\bk}\hat{\bu}(\bk,t)\,,
\end{equation}
where ${\bf v}(\bx,t)$ is the representation of the decimated velocity field in the real space. 
The factors $\gamma_{\bk}$ are chosen to be  either 1 or 0 with the following rule:
\begin{equation}
\label{eq:theta}
\gamma_{\bk} = \cases{
1, & with probability $h_k$  \\
0, & with probability $1-h_k, \, k\equiv|{\bf k}|\,$.}
\end{equation}
Once defined, the set of factors $\gamma_{\bk}$  are kept unchanged, quenched in time.
Moreover, the factors $\gamma_{\bk}$ preserve Hermitian
symmetry $\gamma_{\bk}= \gamma_{-\bk}$ so that ${\cal P}$ is a
self-adjoint operator. 

The NS equations for the Fourier decimated velocity field are then,
\begin{equation}
\label{eq:decimNS}
\partial_t {\bf v} = {\cal P}[- {\bf \nabla}p -
(\bf v \cdot {\bf \nabla}  \bf v )]  + 
  \nu \,\nabla^2 {\bf v} +  {\bf F}\,. 
\end{equation}

We notice that in the above definition of the decimated NSE, the
nonlinear term must be projected on the quenched decimated set, to
constrain the dynamical evolution to evolve on the same set of Fourier
modes at all times. Similarly, the initial condition and the external
forcing must have a support on the same decimated set of Fourier
modes. In the $(L^2)$ norm, $\|{\bf v}\| \propto \int |{\bf v}({\bf
  x})|^2 d^3 x$, the self-adjoint operator ${\cal P}$ commutes with
the gradient and viscous operators. Since ${\cal P}\, {\bf v}= {\bf
  v}$, it then follows that the inviscid invariants of the dynamics
are the same of the original problem in $D=3$, namely energy and
helicity.

\begin{table*}
{\tiny \begin{center}
\begin{tabular}{@{\extracolsep{\fill}} |c|c|c|c|c|c|c|c|}
\hline $N$ & $\nu$ & $\epsilon$ & $\tau_\eta$ & D $(Re_{\lambda})$ & $\alpha$ $(Re_{\lambda})$ & $[\%]$\\
\hline \hline 512 & 0.001 & 0.79 $\pm$ 0.03 & 0.035 $\pm$ 0.002 & $\frac{3 (69); 2.99 (70); 2.98 (70);}{ 2.95 (70); 2.9 (72); 2.8 (74)}$ & ------ & 0; 0.03; 0.06; 0.15; 0.28; 0.46\\
\hline \hline 512 & 0.001 & 0.80 $\pm$ 0.01 & 0.035 $\pm$ 0.001 & ------
  & $\frac{0.97 (72); 0.95 (71); 0.93 (72);}{ 0.9 (73); 0.7 (72); 0.5 (82)}$ & 0.03; 0.05; 0.07; 0.1; 0.3; 0.5\\
\hline \hline 1024 & $8 \times 10^{-4}$ & $1.4 \pm 0.2 $ & $0.023 \pm 0.005$ & $\frac{3 (129); 2.99 (125);}{ 2.9 (131)}$
       & ----- & 0; 0.04; 0.34\\
\hline \hline 1024 & $3 \times 10^{-4}$ & $1.4 \pm 0.2 $ & $0.015 \pm 0.003$ & 2.8 (220)
       & ----- & 0.66\\
\hline \hline 1024 & $8 \times 10^{-4}$ & $1.4 \pm 0.1 $ & $0.023 \pm 0.003$ & ------
       & $\frac{0.99 (126); 0.95 (126);}{ 0.90 (127)}$ & 0.01; 0.05; 0.1\\       
\hline \hline 1024 & $3 \times 10^{-4}$ & $1.3 \pm 0.2 $ & $0.015 \pm 0.003$ & ------
       & 0.60 (205) & 0.4\\       
 \hline
\end{tabular}
\caption{Parameters of our direct numerical simulations: $N$ the
  number of grid points along each spatial direction; $\nu$ the
  kinematic viscosity; $\epsilon$ the mean energy dissipation rate;
  $\tau_\eta \equiv \sqrt{\nu/\varepsilon}$ the Kolmogorov time
  scale. The Reynolds number is estimated as $Re_{\lambda}\equiv
      \sqrt{u_{\rm rms} L /\nu}$, where $u_{\rm rms}$ is the
      root-mean-square value of the velocity, and $L= 2 \pi$ is the
      size of the system; note that at changing the decimation
      importance, the Reynolds number of the flow can also change, due
      to an increase of the total kinetic energy in the system (see
      discussion below). The Reynolds number is given for each run,
      together with the fractal dimension $D$ or the $\alpha$
      values. $D$ is the dimension of the Fourier set for the
  fractally decimated cases. The values $\alpha$ are the probability
  used in the cases of homogeneous decimation and $[\%]$ in the last
  column is the percentage of decimated modes with respect to the non
  decimated case: it is estimated in the wavenumber range between the
  mode zero up to the wavenumber where the energy dissipation spectrum
  peaks.}
\label{sim-details}
\end{center}}
\end{table*}

In this work we adopt two different projectors ${\cal P}$ based on different
definitions of the  $h_{k}$ factors. One is given by a \emph{fractal Fourier}
decimation, first introduced in \cite{frisch2012} as: $$h_k \propto
(k/k_0)^{D-3}\,,\quad\quad \mbox{with}\quad \quad 0< D \le 3\,,$$ where $k_0$
is a small wavenumber here always taken to be 1.  This decimation ensures that
the dynamics is restricted isotropically to a $D$-dimensional Fourier
sub-space. Note that this implies that the velocity field is embedded in a
three-dimensional space, but effectively possesses a set of degrees of freedom
(DOF) inside a sphere of radius $k$ growing as $\#_{DOF}(k) \sim k^{D}$. The
smaller the fractal dimension $D$, the slower is the associated growth of the
DOF. Moreover, the decimation clearly has a larger impact towards the
ultra-violet cutoff, since modes in the high wave number range have a larger
probability to be decimated. Note that this is different from studying NSE in geometries with one compactified dimension, as previously reported \cite{Celani2010}.

The second choice consists of keeping the degree of mode reduction  {\it
homogeneous} in the wave number range: $$h_k = \alpha,\quad \forall k;  \qquad
\mbox{with} \qquad 0 \le \alpha \le 1\,.$$

\subsection{Set-up of the numerical experiments }
\label{sec:numerics}

We performed different series of direct numerical simulations of the
incompressible Navier-Stokes equations in a 
2$\pi$-periodic volume, with a standard pseudo-spectral approach fully
dealiased with the two-thirds rule. Time stepping is done with a
second-order Adams-Bashforth scheme. 

The first set of simulations is done by using $N^3=512^3$ collocation
points. In these runs, a constant energy injection
forcing~\cite{pope_forc,pandit} acting only at large scales, $1 \le
k_{\rm force}\le 2$, is implemented to keep the system in a
statistically steady state. The second set of simulations is done with
$N^3=1024^3$ grid points. A statistically steady, homogeneous and
isotropic turbulent state is maintained by forcing the large scales,
$0.5 \le k_{\rm force}\le 1.5$, of the flow via a second-order
Ornstein-Uhlenbeck process~\cite{sawford}. The choice to adopt a
time-correlated process for the forcing is dictated by the requirement
to enforce the continuity of the acceleration of particles. In all
simulations with $N=1024$, the correlation time-scale of the forcing
is $\sim 10 \tau_{\eta} $ so that small scales are unaffected by the
precise forcing mechanism. For both resolutions, we perform different
sets of DNS for different values of the fractal dimension $D$ or the
decimation percentage $\alpha$. In our study, we define the
    Reynolds number as $Re_{\lambda}\equiv \sqrt{u_{\rm rms} L /\nu}$,
    where $u_{\rm rms}$ is the root-mean-square value of the velocity
    field and $L= 2 \pi$ is the size of the system. In Table
    ~\ref{sim-details}, for each run we report the fractal dimension
    $D$ or the $\alpha$ values, together with the estimated Reynolds
    number.\\

To obtain the Lagrangian statistics, we seeded the flow with tracer
particles. The particles do not react on the flow and do not interact
amongst themselves. The trajectories of individual particles are
described via the equation:
$$\dd{\Xp}{t} = {\bf v}(\Xp(t),t)\,,$$ and are integrated by using a
trilinear or B-spline 6th order interpolation scheme ~\cite{NumRec},
to obtain the fluid velocity, ${\bf v}(\Xp(t),t)$, at the particle
position. We note that Galerkin truncation or decimation operators
destroy the Lagrangian structure of the NS dynamics.  However, it is
clear that studying the Lagrangian dynamics in turbulent decimated
flows is always possible, and, as we will see, it is an important
piece of information when dealing with the nature of intermittency in
hydrodynamical turbulence. 
\begin{figure*}
\begin{center}
\includegraphics[width=0.49\linewidth]{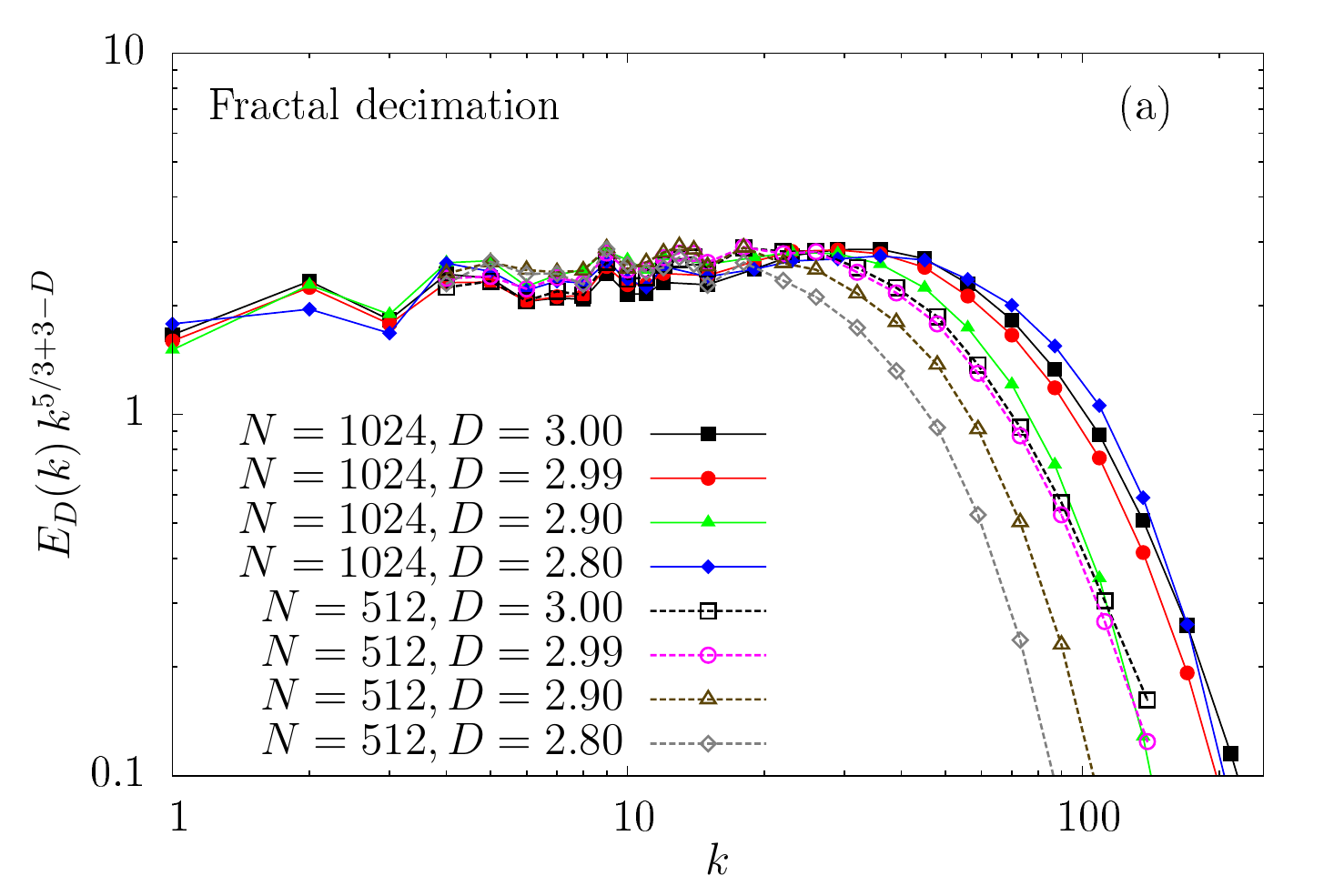}
\includegraphics[width=0.49\linewidth]{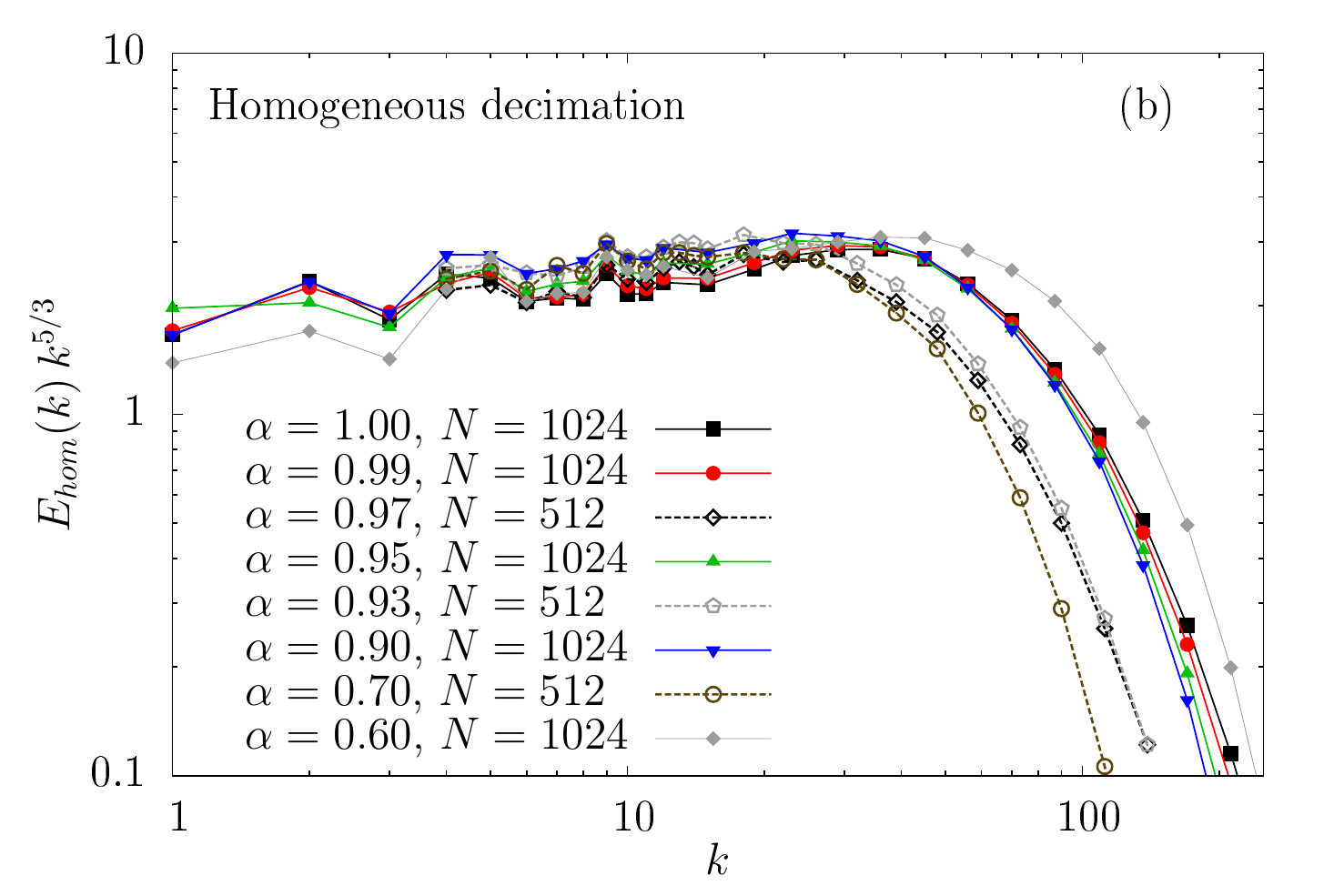}
\end{center}
\caption{Compensated kinetic energy spectra (a) $k^{5/3 + D - 3}E(k)$
  for fractal and (b) $k^{5/3}E(k)$ for homogeneous decimation. The
  spectra have been shifted to obtain a collapse onto a single curve,
  for ease in visualization.}
\label{fig:spectra}
\end{figure*}

\section{Results}
\label{sec:results}
\subsection{Spectra}
As shown in \cite{PRL2015}, fractal decimation induces a correction,
$\propto k^{3-D}$, for the power law scaling of the kinetic energy
spectrum:
\begin{equation}
\label{eq:spectrum}
E_{D}(k) \sim k^{3-D}\,k^{-5/3}\,,
\end{equation}
where the factor $\sim k^{-5/3}$ is the K41 spectrum predicted by
Kolmogorov in $1941$ theory and valid for the $D=3$ original problem
(we neglect intermittent corrections). The derivation of this result
can be found in \cite{PRL2015}: it is based on the empirical
observation that the energy flux, $\epsilon$, remains constant in the
inertial range of scales and for all fractal dimensions. In order to
keep a constant flux across all scales, with less and less modes, the
spectrum must acquire a power-law correction. Note that the extra
power-law correction induced by the fractal decimation introduces new
contributions in the Eulerian domain, leading to a complex
superposition of scaling properties as shown in
\cite{PRL2015,EPJE2016,PRE2016}. Furthermore, this makes it even more
difficult to interpret Lagrangian statistics starting from the
Eulerian phenomenology. \\Since for the homogeneous case the decimation
probability is constant and independent of $k$, these difficulties are
absent and we expect a K41 spectra for all $\alpha$:
\begin{equation}
\label{eq:spectrum-hom}
E_{hom}(k) \sim  k^{-5/3}.
\end{equation}
 In Fig. {\ref{fig:spectra} we confirm the two predictions
   (\ref{eq:spectrum})-(\ref{eq:spectrum-hom}) by showing the
   compensated energy spectra for the case of (a) fractal ($E_D(k)
   k^{D-3 + 5/3}$) and (b) homogeneous ($E_{hom}(k)
   k^{5/3}$)decimations. The curves all collapse for all the values of
   $D$ and $\alpha$.\\ Before concluding this section, we comment
       that the power-law correction of the spectrum exponent for the
       fractal cases is such that for $D=7/3$, the spectrum becomes
       divergent in the region of large $k$. Numerically, the
       investigation of the system at such low fractal dimension is
       critical, since e.g. at resolution $N=1024$ about less than 1\%
       of the modes would survive. The actual behavior of the system
       at dimensions close to $7/3$ remains an open question.
\begin{figure}[h!]
\begin{center}
\includegraphics[width=1.\columnwidth]{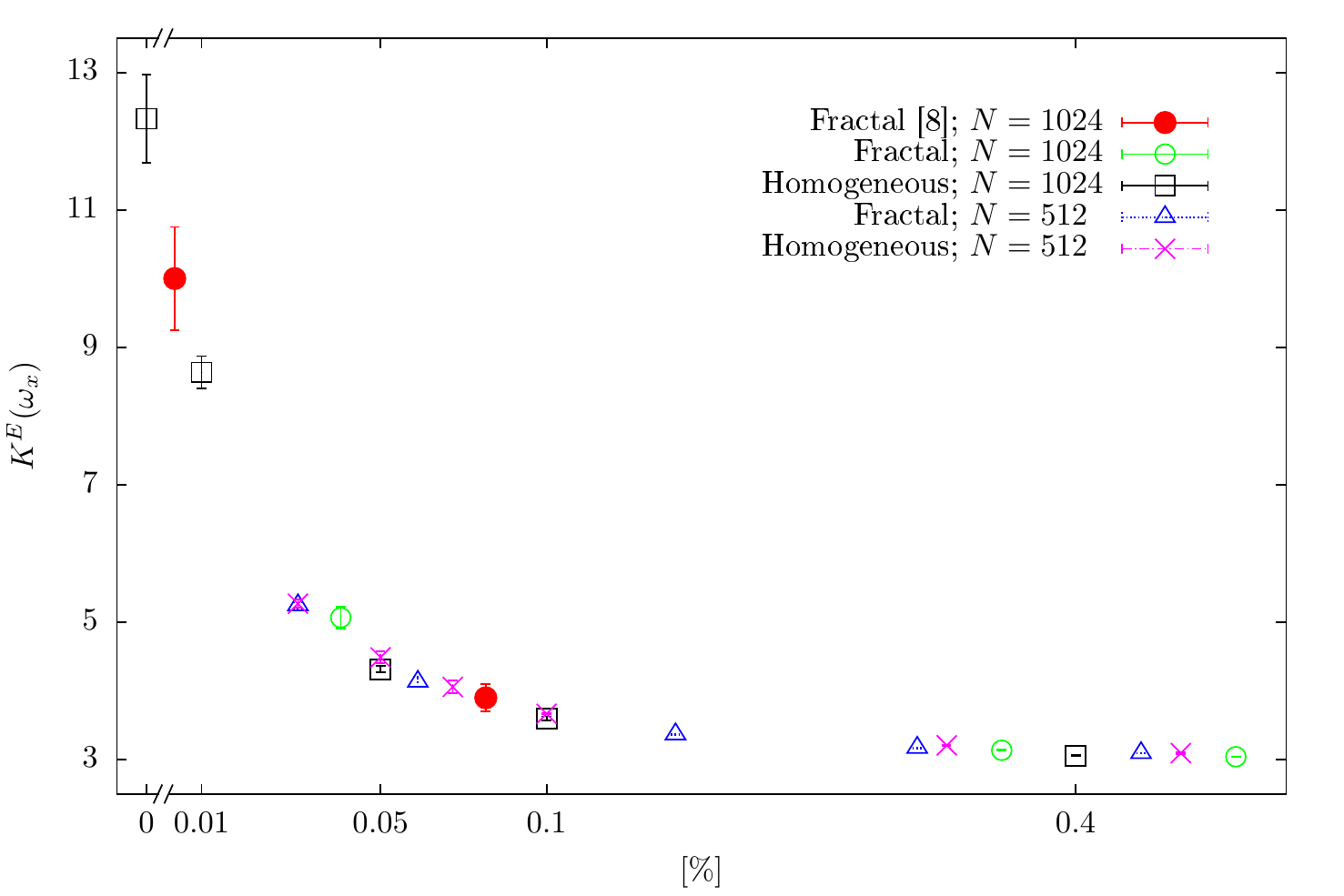}
\end{center}
\caption{The kurtosis of the x-component of the vorticity field
  $K^E(w_x)$ vs the percentage of decimated Fourier modes, $[\%]$. The
  explicit values of the percentage of decimation and the
  corresponding fractal dimension $D$ or the probability of
  homogeneous decimation $\alpha$ are listed in
  Table~\ref{sim-details}.}
\label{fig:vort-flatness}
\end{figure}

\subsection{Higher-Order Eulerian Statistics}
\label{subsec:Eulerian}
In three-dimensional turbulence, the distribution of the spatial
derivatives of the velocity field has a strongly non-Gaussian
behavior. A measure of this is given by the kurtosis of one component
of the vorticity field (assuming small-scales isotropy):
$$K^E(w_x)= \frac{\langle \omega_x^4 \rangle}{\langle
  \omega_x^2\rangle^2}\,\, . $$ A Gaussian distribution is
characterized by $K^E=3$. It is known that in three-dimensional
turbulence, the kurtosis is larger than 3 and grows as a function of
the Reynolds number, indicating that the flow is becoming more and
more intermittent.  It was previously observed \cite{PRL2015,EPJE2016}
that when fractal decimation is applied, the kurtosis approaches the
Gaussian value with decreasing $D$.  In Figure \ref{fig:vort-flatness}
we show the value of the kurtosis $K^E(w_x)$ as a function of the
percentage of removed modes $[\%]$ (defined below and listed in
Table~\ref{sim-details}) for both fractal and homogeneous decimations
as well as for the different Reynolds numbers.  To make the comparison
between the fractal and homogeneous protocols meaningful, we use the
percentage of mode reduction in the fractal case measured as the
percentage of modes removed up to $k_{\rm peak}$, i.e., up to the
wavenumber where the dissipation energy spectrum $k^2 E_D(k)$
peaks. For homogeneous decimation, the percentage of removed modes
$[\%]$ is simply $1 - \alpha$.  Our results show that not only does
the homogeneous decimation cause a suppression of intermittency, but
the effect takes place with the same dependence on the percentage of
removed modes as measured in the fractal case. Moreover, we see that
our data from these sets of simulations are in agreement with data
from \cite{PRL2015} where a different forcing was used. To summarize,
turbulent decimated systems show a unique tendency towards a
quasi-Gaussian statistics, independent of the decimation protocol.

The suppression of spatial intermittency under decimation leads us to
the main question of this paper: what happens to the Lagrangian
dynamics when small-scale structures responsible for the vortex
stretching are largely modified \cite{EPJE2016}, if not destroyed? To
answer this question we stick to the homogeneous decimation case in
order to avoid the further complication induced by the power-law
correction present in the velocity scaling in the fractal case.

\subsection{Lagrangian Statistics}
\label{subsec:Lagrangian}
In this section, we analyze the statistical behavior of tracer
particles in decimated flows. In order to do that it is useful to
define the order-$p$ Lagrangian structure function:
\begin{equation}
S^L_p(\tau) \equiv \sum_i\, \langle [\delta_{\tau} v_i]^p\rangle \sim \, \tau^{\zeta_p^L}\,,
\label{eq:SFlag}
\end{equation}
where $ \delta_\tau v_i = v_i(\Xp(t+\tau),t+\tau) - v_i(\Xp(t),t)$ and
the sum is over the three components of the velocity field (assuming
isotropy).  As for the Eulerian case, we quantify deviations from
Gaussian statistics at changing time lags by defining the Lagrangian
kurtosis:
\begin{equation}
K^L(\tau) =\frac{S^L_4(\tau)}{(S^L_2(\tau))^2}.
\end{equation}
$K^L(\tau)$ evaluated at time increment $\tau = \tau_\eta$ is plotted
in Fig.~\ref{fig:kurtosis-LAG}. It shows a strong dependence on the
number of DOF, similar to what happens for the Eulerian case.  The
dependence on the Reynolds number of the flow is weak as we find that
the measurements done in the DNS with $N=512$ or $N=1024$ exhibit the
same behavior. Notice that in Figure ~\ref{fig:kurtosis-LAG}, the
    value of the kurtosis for the smallest time scale is very close to
    the one obtained by looking at the acceleration of the
    particles.

To quantify the Lagrangian properties at all time lags $\tau$, we show
in the inset of the same figure the kurtosis of tracers for all
$\tau$. It is clear that there is a rapid reduction of intermittency,
as it was reported in \cite{PRL2015}, just like the corresponding
Eulerian measurements made for spatial increments.  Since Lagrangian
statistics are known to be more intermittent than their Eulerian
counterparts (as quantified by the deviations from the dimensional
scaling), this result is even more interesting. This is because it
shows how a nominally small decimation ($\alpha=0.95$) is responsible
for a decrease of about $85\%$ of the kurtosis at small $\tau$. We
attribute such a large reduction to the strong modification of intense
vortical structures as reported in a previous study
\cite{EPJE2016}. Moreover, the quick recovery of quasi-Gaussian
statistics by increasing the degree of decimation is, for this
observable and for the Reynolds numbers investigated here, almost
independent of the Reynolds number. It is also noteworthy that the
results are independent of the particular type of large-scale forcing
since the form of forcing for $N= 512$ differs from the case of
$N=1024$.
\begin{figure}[h!]
\begin{center}
\includegraphics[width=1.\columnwidth]{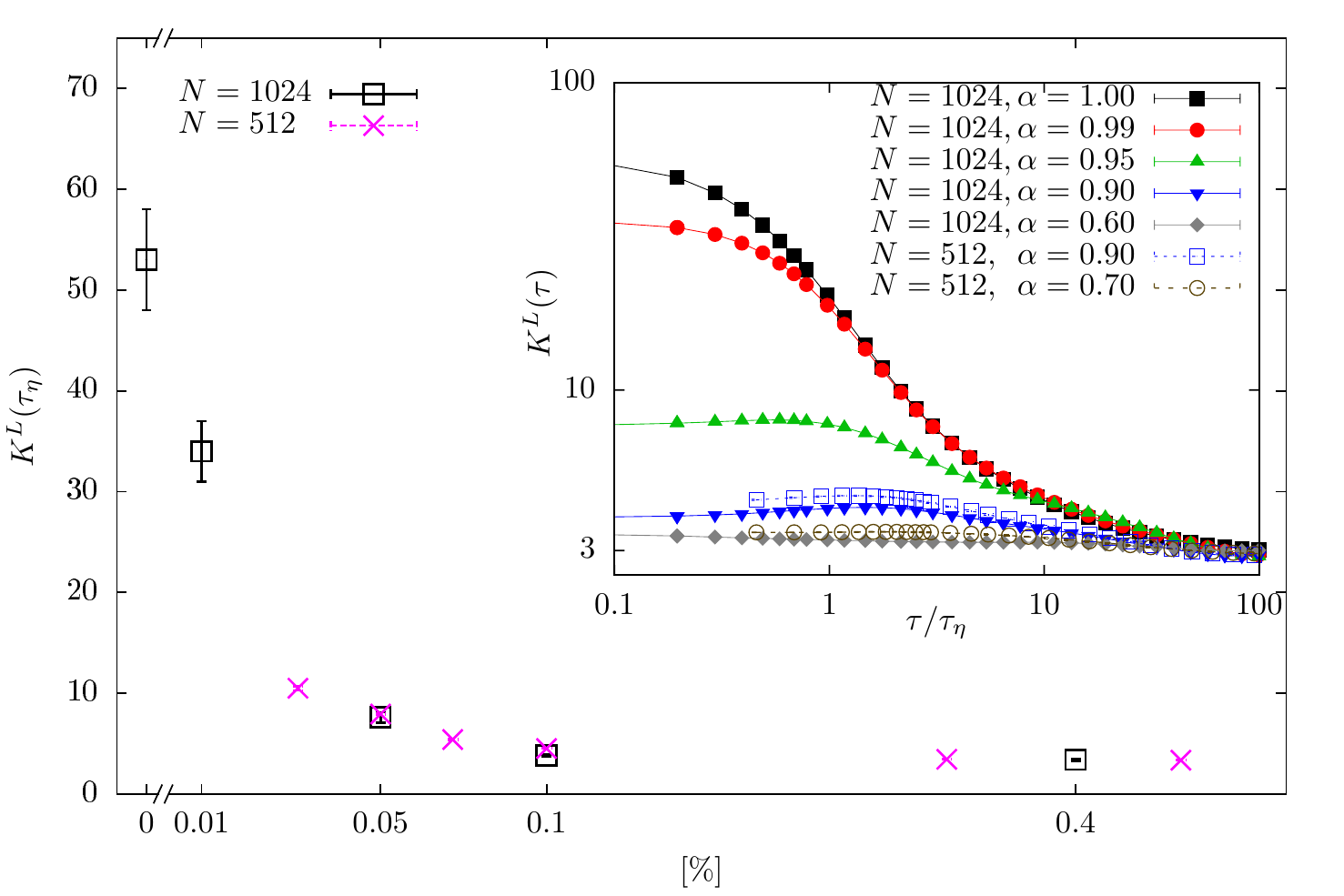}
\end{center}
\caption{The kurtosis $K^L(\tau)$ of the Lagrangian structure function
  measured at the Kolmogorov time scale $\tau = \tau_{\eta}$ as a
  function of the percentage of decimated Fourier modes $[\%]$. In the
  inset we show representative plots of $K^L(\tau)$ vs $\tau$ for some
  values of $\alpha$}
\label{fig:kurtosis-LAG}
\end{figure}

\begin{figure}[h!]
\begin{center}  
\includegraphics[width=1.\columnwidth]{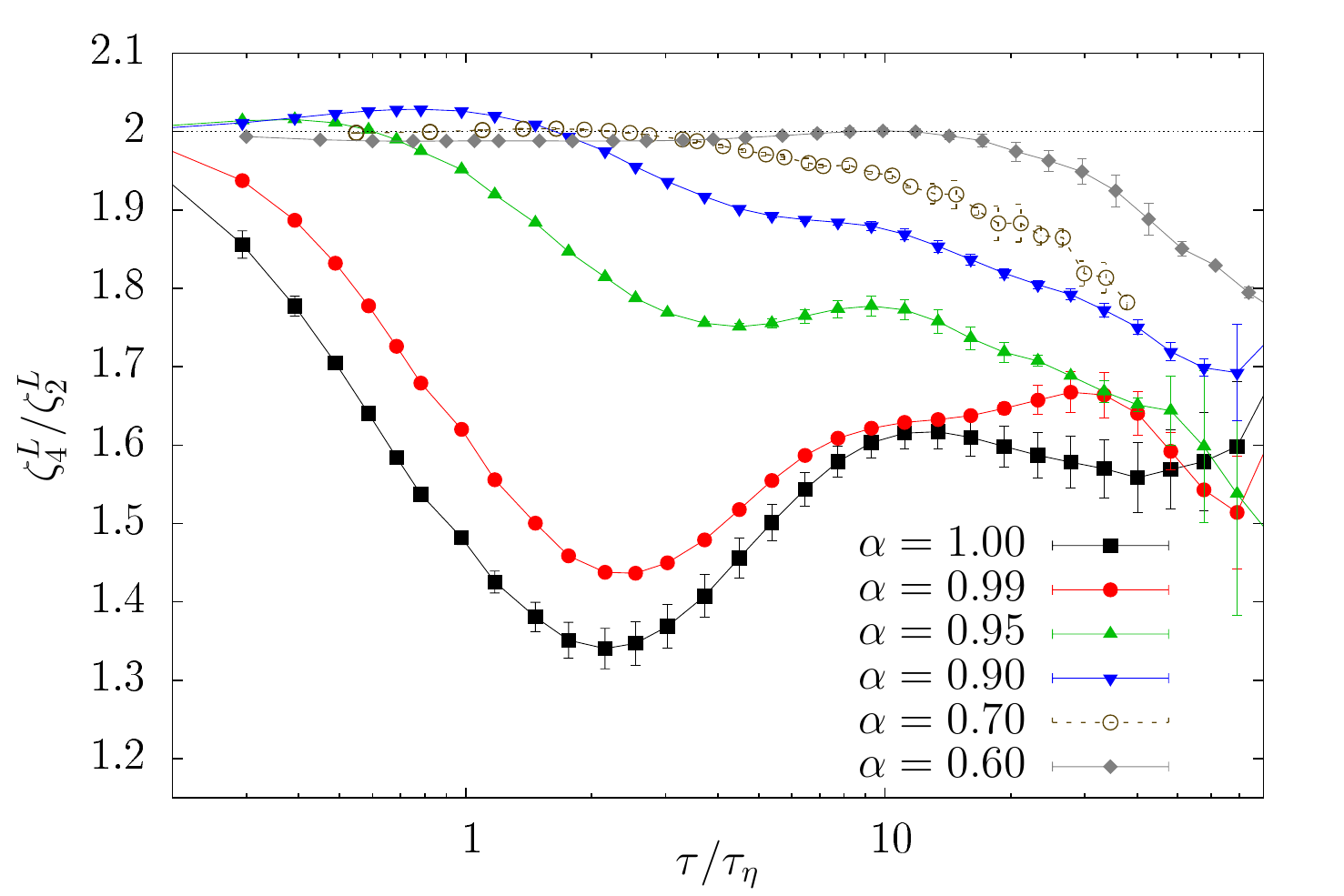}
\end{center}
\caption{Log-lin plot of the local slopes of the scaling exponent
  $\zeta_4^L/\zeta_2^L$ of the Lagrangian structure functions,
  averaged over the three velocity components, vs $\tau/\tau_{\eta}$
  for some representative value of $\alpha$ from the simulations at
  $N=1024$.}
\label{fig:tracers-ess}
\end{figure}

To have a deeper understanding of the Lagrangian scaling, in
Figure~\ref{fig:tracers-ess}, we plot the {\it local slopes} by using
Extended Self Similarity (ESS) \cite{PRL2008,ess1,ess2}, i.e., the
logarithmic derivative of the fourth order Lagrangian structure
function, $S^{(4)}_{L}(\tau)$, versus the second order one,
$S^{(2)}_{L}(\tau)$, which gives:
\begin{equation}
\frac{d \log S^L_4(\tau)}{d \log S^L_2(\tau)}=\frac{\zeta^L_4}{\zeta^L_2}\,.
\label{eq:z4z2}
\end{equation}
Let us notice that the above quantity is a direct {\it scale-by-scale}
measurement of the local scaling properties and does not need any
fitting procedure. A scale-independent behavior of one moment against
the second-order one would result in a constant value for the left
hand side of (\ref{eq:z4z2}). We recall that in the absence of
intermittency, these curves should be constant across the time lags
with $\zeta^L_4/\zeta^L_2 = 2$; while this relation is always well
verified for the smooth dissipative scales, a non trivial behavior
appears in the inertial range pointing out the intermittent feature of
the original system. A few important observations should be
made. First, in the time range from $1$ to $10\,\tau_{\eta}$, the
strong deviation observed in the local slope of the standard $D=3$
case and attributed to events of tracer trapping in intense vortex
filaments \cite{POFtrapping2005} rapidly disappears as soon as the
mode reduction is applied. Second, we observe that in the inertial
range (where the $D=3$ local exponents develop a plateaux) the scaling
for the decimated cases is much poorer, i.e. the local-slopes are no
longer constant.  Finally, independent of the existence of a pure
scaling behavior, we observe that the intermittent correction is also
reduced as the percentage of removed modes is increases, and it almost
vanishes, reaching the dimensional value $2$ for almost all $\tau$,
already at $\alpha=0.6$, corresponding to $40\%$ of the modes
decimated. These observations are valid for all the Reynolds numbers
here explored.
\begin{figure*}
\begin{center}
\includegraphics[width=0.49\linewidth]{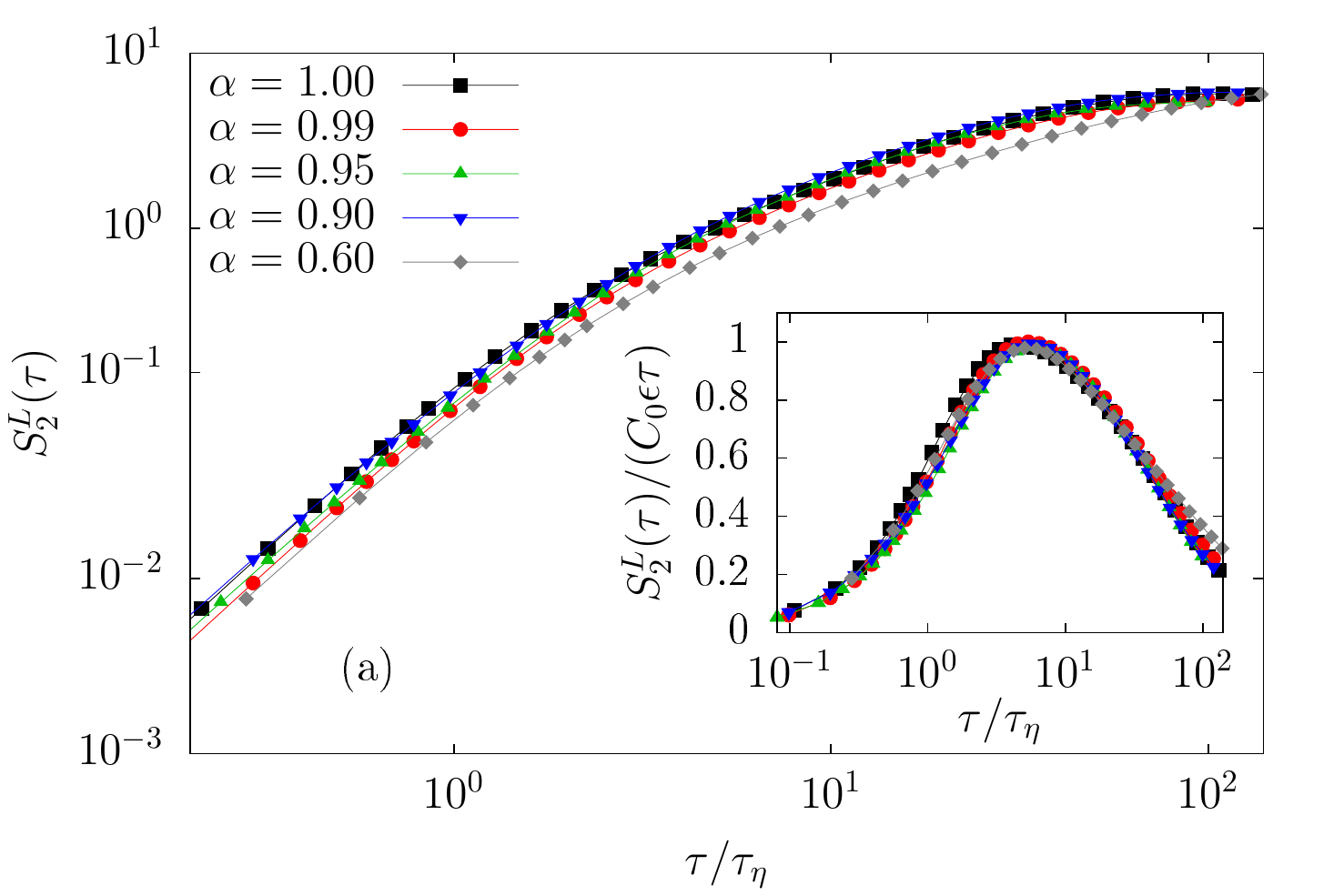}
\includegraphics[width=0.49\linewidth]{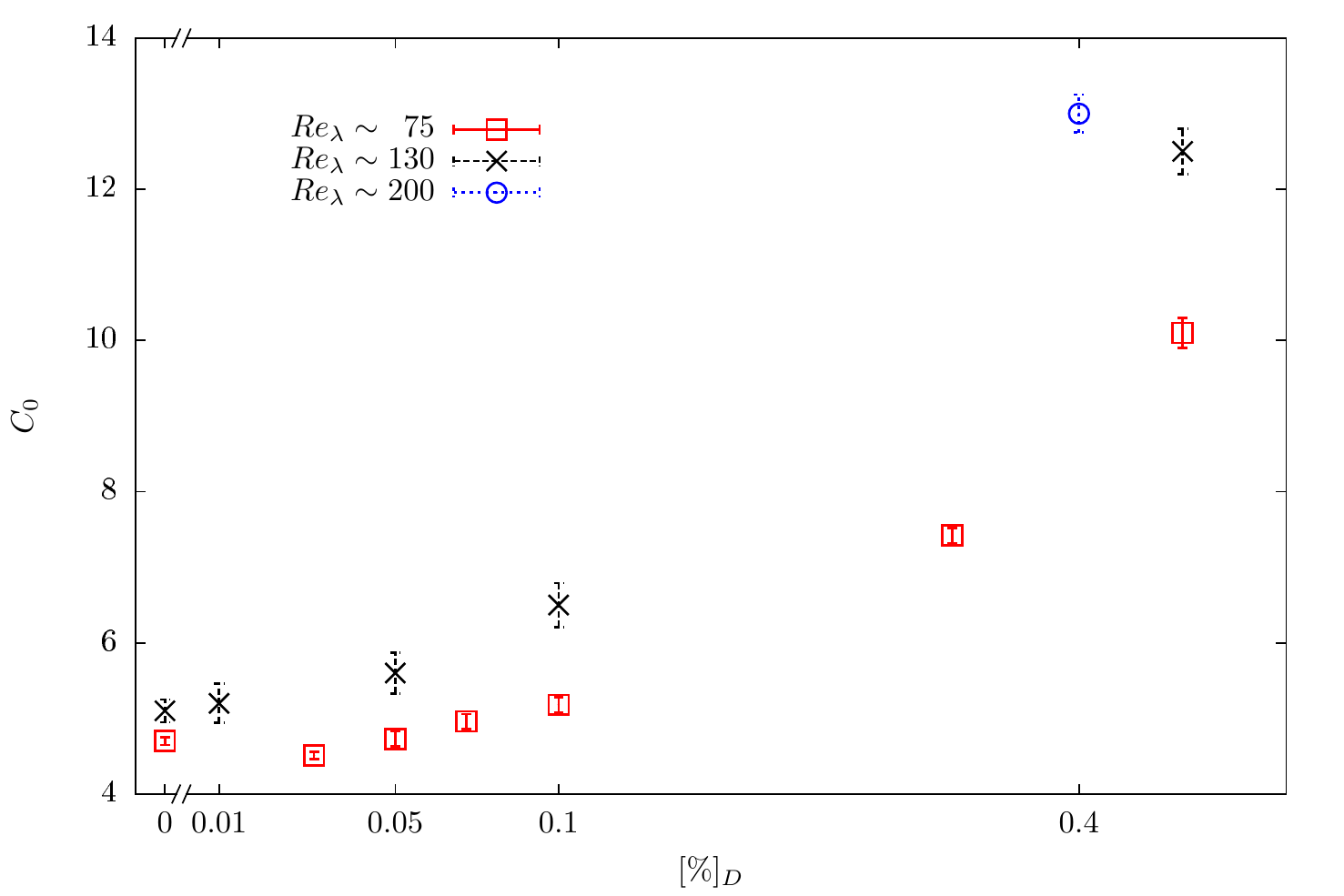}
\end{center}
\caption{(a) Log-log plot of the second-order Lagrangian structure
  function $S^L_2(\tau)$ vs $\tau/\tau_\eta$, averaged over the three
  velocity components, for some representative values of $\alpha$. In
  the inset we show the linearly compensated second-order Lagrangian
  structure function, $S^L_2(\tau)/(C_0\,\epsilon \tau)$; the
  dimensionless constant $C_0$ is chosen such that the peak of the
  compensated structure function is 1. In (b) we show the
      variation of $C_0$ as a function of $[\%]$ for the homogeneously
      decimated cases, and for runs at comparable Reynolds numbers:
      the upper curve is for $Re_{\lambda} \simeq 130$ ($\times$),
      while the lower curve is for $Re_{\lambda} \simeq 75$
      (squares). The isolated circle point is for the numerical
      simulation at $N=1024$, viscosity $nu= 3.e-4$ and $\alpha=0.6$,
      for which the estmated Reynolds number is higher than the other
      cases.}
\label{fig:s2}
\end{figure*}


\section{Connecting Eulerian and Lagrangian Statistics in Decimated Flows}
\label{sec:bridge}
An important open point in literature is connected to the relation
between Eulerian and Lagrangian statistics
\cite{Borgas1993,PRL2008,Boff2002,Chevillard2003,Benzi2010,HKFG2009,Falk2012,Leveque2014}.
The two ensembles must of course be correlated.  Let us introduce the
order-$p$ Eulerian structure function in a manner analogous to the
definition of the Lagrangian structure function. For the longitudinal
velocity increments $\delta_r v = [{\bf v}({\bf x} + {\bf r}) - {\bf
    v}({\bf x})]\cdot {\hat r}$, the longitudinal Eulerian structure
function can be written as
\begin{equation}
S^E_p(r) \equiv \langle [\delta_r v]^p\rangle \sim r^{\zeta^E_p}.
\label{eq:SFeul}
\end{equation}
Similarly, one could have introduced transverse Eulerian structure
functions, based on transverse increments \cite{frischbook}.
Dimensional predictions based on the idea that in the inertial range
everything is driven by the energy transfer rate, $\epsilon$, puts
strong constraints on the possible functional dependencies of Eulerian
and Lagrangian structure functions. For example, dimensional
predictions give $\zeta_p^E=p/3$ and $\zeta_p^L = p/2$.  It is also
known that in the presence of intermittent corrections, where
$\zeta_p^E \ne p/3$ and $\zeta_p^L \ne p/2$, the two sets of exponents
are well explained by a {\it bridge relation}
\cite{Borgas1993,PRL2004,Schmitt2006,Boff2002}. The idea is to connect
the spatial and temporal fluctuations over increment $r$ and $\tau$ by
\begin{equation}
\delta_{\tau} v \sim \delta_{r}v; \qquad \qquad \tau \sim r/\delta_r v\,.
\label{eq:MF}
\end{equation}
Applying the usual multifractal formalism, is then possible to show
that the following relation holds
\cite{Borgas1993,PRL2004,Schmitt2006,Boff2002}:
\begin{eqnarray}
\label{eq:bridge12}
&& \zeta^L_p=\zeta^E_n\,,\\
&& p\,=n\, - \zeta^E_n\,.
\label{eq:bridge}
\end{eqnarray}
It is important to notice that the above relation is consistent with
the dimensional phenomenology. Moreover, considering that we have the
exact Eulerian result $\zeta_3^E=1$, the second order Lagrangian
structure function must scale linearly according to (\ref{eq:bridge}),
$\zeta_L^2=1$:
 \begin{equation}
\label{eq:s2lag}
S^L_2(\tau) = C_0 \epsilon \tau\,.
 \end{equation}
Different scaling properties for three-dimensional Lagrangian
turbulence have also been proposed, as discussed in
\cite{Falk2012}. The question of whether this bridge relation is exact
or a very good first-order approximation is still open.  Since, even
under decimation, one can prove that $\zeta_3^E=1$, it is important to
check whether the above prediction (\ref{eq:s2lag}) still holds
(empirically) under the application of homogeneous decimation.  In
Figure~\ref{fig:s2}(a), we plot the second order moment of velocity
increments, averaged over the three field components, for the
homogeneously decimated runs at $N=1024$. We note that all curves
exhibit a similar behavior, which also means that there is no evident
difference between data of $D=3$ standard turbulence, and the data
from the decimated cases, in agreement with (\ref{eq:s2lag}).

In the inset of the same figure we also plot the compensated curves,
$S_2^E(\tau)/(C_0\epsilon \tau)$ versus $\tau$. Looking at the
compensated plots, we can observe better the agreement among $D=3$ and
decimated turbulence. Such a good overlap of the different curves is
obtained by accurately fixing the values of two parameters. First, the
Kolmogorov time scale $\tau_{\eta}$ is varied within the error bars
given in Table~\ref{sim-details} to obtain an optimal horizontal
shift. Second, the coefficient $C_0$ is also changed in order to fix
the peak of the correlation at $1$. The value of normalization
    constant $C_0$ has been already examined in previous works and it
    is known to depend on the Reynolds number of the flow, see
    e.g. \cite{yeung_rev,yeung_recent}. For high Reynolds numbers in
    three-dimensional turbulence, it is estimated that $C_0$ lies in
    the range $6-7$. In our simulations, we have measured for $D=3$
    the value $C_0=5.2$ for $N=1024$, which is in agreement with the
    previous measurements at the same Reynolds number. The measured
    behavior of $C_0$ as a function of the percentage of removed modes
    $[\%]$ is interesting. In panel (b) of Figure~\ref{fig:s2}, we see
    that it grows as the reduction of the DOF in the system
    increases. This result is in agreement with the observation first
    reported in \cite{frisch2012} that, at increasing decimation, a
    less efficient energy transfer towards small-scale leads to a
    growth of total kinetic energy due to an accumulation at the
    largest scales of the system. We also note that for any given
    value of the percentage of modes decimated, $C_0$ is slightly
    larger for the homogeneous runs with higher Reynolds number. We
    remark that $C_0$ is known to be strongly sensitive to the
    underlying Eulerian flow realization, since for example in
    two-dimensional (undecimated) turbulence in the inverse cascade
    regime \cite{JOT2013}, it can become as large as 40-50 depending
    on the inertial range extension.\\ The scaling properties of
$S^L_2(\tau)$ show that the bridge relation is robust under mode
reduction, at least for those observables that are not affected by
intermittency. We now ask the same question for higher order moments,
where intermittency play a major role and strongly depends on the
degree of mode reduction.  In Figure~\ref{fig:bridge}, we test on the
numerical data the {\it bridge-relation} (\ref{eq:bridge}) for the
Lagrangian scaling exponent $\zeta^L_4/\zeta^L_2$. To do this, we
first need to estimate from the Eulerian data a functional form for
the curve of the scaling exponents, $\zeta^E_p$ of the structure
functions for different values of $p$. We accomplish this by repeating
the procedure illustrated e.g. in \cite{PRL2008}, by using the
multifractal model and a log-Poisson distribution for the singularity
spectrum. Details are not repeated here for the sake of brevity. \\The
shaded area around each curve represents our uncertainty on the local
Lagrangian scaling exponents, based on the Eulerian ones: indeed in
the Eulerian framework, scaling exponents are not defined uniquely
since longitudinal and transverse moments are observed to scale
differently, see e.g., \cite{PhysD2008}. The shaded area is associated
to the different sets of Lagrangian scaling exponents that we can
obtain considering either the longitudinal or the transverse moments
scaling in the Eulerian framework, and hence in the relations (\ref{eq:bridge12}) and (\ref{eq:bridge}).  The agreement is remarkable.

To summarize, we have tested, a non-linear set of relations bridging
Eulerian scaling exponents to Lagrangian ones. These transformations
are based on the idea that statistical relations among velocity
singular fluctuations do survive decimation protocols, and can then be
used even when the detailed form of the Navier-Stokes equations is
modified by a strong reduction of the degrees of freedom.

\begin{figure}
\begin{center}
\includegraphics[width=1.0\columnwidth]{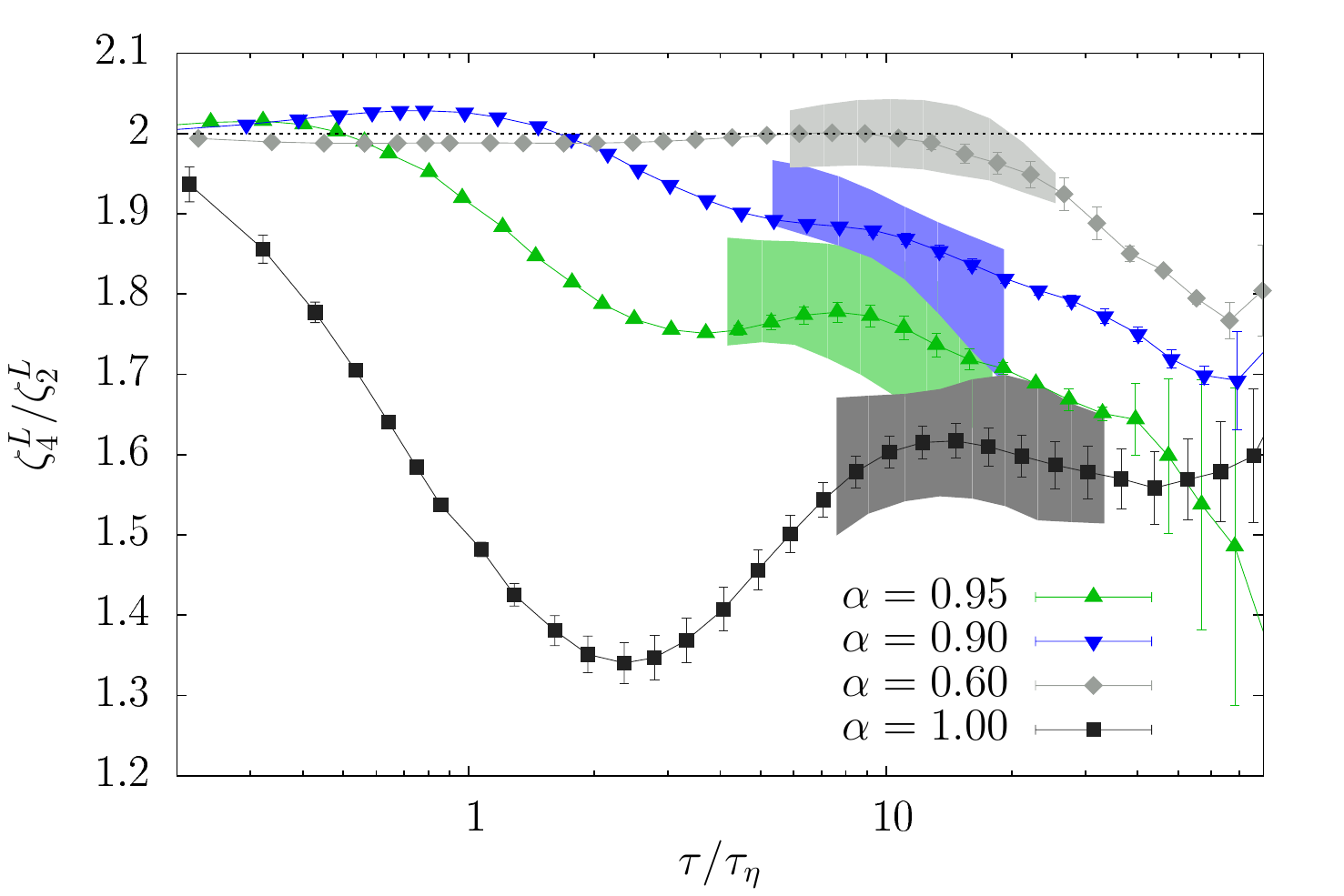}
\end{center}
\caption{Log-lin plot of the local slopes of the scaling exponent
  $\zeta_4^L/\zeta_2^L$ of the Lagrangian structure functions,
  averaged over the three velocity components, vs the time lag
  $\tau/\tau_{\eta},$ for some representative values of $\alpha$ from
  simulations with higher Reynolds number. The shaded region shows the
  inertial range prediction from the multifractal model by using the
  Eulerian longitudinal and transversal structure functions. The
  straight dashed line correspond to the dimensional non-intermittent
  value $\zeta_4^L = 2 \zeta_L^2$. The overlap of the shaded regions
  with the corresponding data from our numerical simulations is a
  confirmation of the validity of the bridge relations (see text),
  connecting the Eulerian and Lagrangian scaling exponents.}
\label{fig:bridge}
\end{figure}

\section{Conclusions}
\label{sec:conclusions}
We have performed a series of direct numerical simulations of the
three-dimensional NSE under Fourier mode reduction.  Projection on a
restricted set of Fourier modes has been largely explored in the past,
starting from the pioneering work of Lee and Hopf to study the Euler
equations with the tools of equilibrium statistical mechanics
\cite{Kr1975,cichowlas,Frisch2008,ray11} or to search for flux-less
solutions with scaling properties close to the Kolmogorov $-5/3$
spectrum in the inverse cascade regime \cite{frisch2012,lvov}.  Here,
we have shown that Fourier mode reduction also offers a unique
opportunity to change the degree of intermittency of the NSE and thus
to study its robustness under a wide spectrum of different
perturbations. Fourier mode reduction has a singular effect on the
dynamics: a weak removal of modes strongly modifies the scaling
properties of turbulent flows.

In this study, we have applied to the original three-dimensional
problem two decimation protocols, where the degree of mode reduction
is changed continuously through different control parameters. In both
cases the resulting dynamics preserves the inviscid conservation
properties and all symmetries of the original problem. Fractal
decimation constraints the set of Fourier modes to live on a fractal
set, with the high-wavenumber degrees of freedom having a larger
probability to be decimated, leading to a larger and larger weight of
non-local Fourier interactions by decreasing $D$. Moreover, fractal
decimation modifies the scaling exponent of the kinetic energy
spectrum, thus introducing a complex superposition of scaling
behaviors in the Eulerian domain. Homogeneous decimation removes
degrees of freedom with the same percentage from large to small
scales, without introducing new scaling properties and keeping the
same statistical weight of local and non-local triadic interactions in
the non-linear evolution.  We have first shown that both protocols
reduce intermittency with the same dependence on the number of DOF in
the system in the Eulerian frame, and at the two Reynolds numbers here
investigated. Some runs have also been repeated by keeping all
parameters unchanged, expect for the stochastic realization of the
decimation mask to check that the statistical measures that we report
are indeed independent of the precise quenched realization of the DOF
reduction.

Concerning the Lagrangian statistics, we have shown that homogeneous
decimation leads to a quick reduction of high-frequency intermittency
too, as measured by the kurtosis of the Lagrangian structure functions
at the Kolmogorov time scale. This reduction of intermittency is
accompanied by a quick increase of the $C_0$ constant in the second
order Lagrangian structure function, indicating a possible singular
behavior in the high decimated regime at large Reynolds numbers. It is
important to recognise that in the limit of infinite Reynolds number,
the fractal and homogeneous decimation protocols will probably lead to
very different asymptotics. This is because, as the Reynolds number
increases, smaller and smaller scales appear which will be decimated
with a larger and larger probability for the fractal decimation
protocol or with a constant probability for the homogeneous case.

Interestingly, in spite of the strong sensitivity of intermittency on
the degree of mode reduction, the two sets of Lagrangian and Eulerian
structure functions remain well-described by a phenomenological {\it
  bridge-relation}, which connects the degrees of intermittency in the
two set of measurements. Besides the previous findings, the outcome of
the present work can be seen as an attempt to characterize the
statistical properties of the NSE when restricted to a reduced set of
modes and before applying a {\it sub-grid} closure for the removed
DOF. This Large-Eddy-Simulation program is typically implemented by
applying a sharp cutoff at $k_c$ in Fourier space for all wavenumbers
with $k >k_c$. Here, the cutoff is still sharp (we apply a projector)
but the grid is diffused among the whole Fourier space, keeping memory
of all scales and frequencies in the system.  This might be crucial to
further improve the modelling of the evolution of particles in
turbulent flows. In particular, the impact of fractal or
    homogeneous mode reductions on the Lagrangian and Eulerian
    statistics can be seen as a first step toward the development of
    models for the removed degrees-of-freedom, which is the ultimate
    goal of any Large-Eddy-Simulation. The strong sensitivity of
    intermittency to the degree of mode reduction is a clear
    indication that this is a delicate issue that needs to be
    investigated with care.

\section*{Acknowledgments} We acknowledge useful discussions with Roberto
Benzi. We acknowledge support from the COST Action MP1305, supported
by COST (European Cooperation in Science and Technology). MB and LB
acknowledge funding from the European Research Council under the
European Union's Seventh Framework Programme, ERC Grant Agreement No
339032. SSR acknowledges the support of the Indo-French Center for
Applied Mathematics (IFCAM), the AIRBUS Group Corporate Foundation
Chair in Mathematics of Complex Systems established in ICTS, the DST
(India) project ECR/2015/000361, and discussions with Jayanta
K. Bhattacharjee.  AB acknowledges support from the Knut and Alice
Wallenberg Foundation under the project ‘Bottlenecks for particle
growth in turbulent aerosols’ (Dnr. KAW 2014.0048).  We acknowledge
that numerical simulations were partly performed at CINECA, within the
INFN allocation for project FIELDTURB and the PRACE grant
N. Pra092256, and on {\it Mowgli} at the ICTS-TIFR, Bangalore, India.

\end{document}